\documentclass[11pt]{article}
\usepackage[top=50mm, bottom=50mm, left=50mm, right=50mm]{geometry}
\usepackage[utf8]{inputenc}
\usepackage{siunitx}
\usepackage{lineno}
\usepackage{amssymb}
\usepackage{amsmath}
\usepackage{amsthm}
\usepackage{epsfig}
\usepackage{graphicx}
\usepackage{graphics}
\usepackage{float}
\usepackage{multirow}
\usepackage{color}
\usepackage{lineno}
\usepackage{fullpage}
\usepackage[normalem]{ulem} 
\usepackage{makeidx}
\usepackage{xspace}
\usepackage{wrapfig}
\usepackage[capitalise]{cleveref}
\usepackage{subcaption}
\usepackage[version=4]{mhchem}
\usepackage[font=small]{caption}
\usepackage{url}

\providecommand{\keywords}[1]
{
  \small	
  \textbf{\textit{Keywords---}} #1
}

\usepackage{authblk}

\title{Controlling the rheo-electric properties of graphite/carbon black suspensions by `flow switching'}
\author[1,2,*]{Thomas Larsen}
\author[3]{John R. Royer}
\author[3]{Fraser H. J. Laidlaw}
\author[3]{Wilson C. K. Poon}
\author[2]{Tom Larsen}
\author[2]{Søren J. Andreasen}
\author[1]{Jesper de C. Christiansen}
\affil[1]{Department of Materials and Production, Aalborg University, Denmark}
\affil[2]{Advent Technologies A/S, Denmark}
\affil[3]{School of Physics and Astronomy, The University of Edinburgh, Scotland}
\affil[*]{Corresponding author: Thomas Larsen, larsenn.thomas@gmail.com}

\date{\today}

\usepackage{setspace}
\usepackage[sorting=none, style=nature]{biblatex}
\bibliography{bibliography} 

\begin{document}

\maketitle

\begin{abstract}

The ability to manipulate rheological and electrical properties of colloidal carbon black gels makes them attractive in composites for energy applications such as batteries and fuel cells, where they conduct electricity and prevent sedimentation of `granular' active components. While it is commonly assumed that granular fillers have a simple additive effect on the composite properties, new phenomena can emerge unexpectedly, with some composites exhibiting a unique rheological bi-stability between solid-like and liquid-like states. Here we report such bi-stability in suspensions of non-Brownian graphite and colloidal carbon black in oil, a model system to mimic composite suspensions for energy applications. Steady shear below a critical stress elicits a transition to a persistent mechanically weak and poorly conducting state, which must be `rejuvenated' using high-stress shear to recover a stronger, high-conductivity state. Our findings highlight the highly tunable nature of binary granular/gel composite suspensions, and present new possibilities for optimizing mixing and processing conditions for Li-ion battery slurries.

\end{abstract}

\keywords{Colloidal, bipolar plate, Li-ion, slurry, conductivity, rheology}

\newpage

\section{Introduction}

As global energy demand increases \cite{chart:IEA,article:Barrett2022}, sustainable storage and conversion technologies (solar cells, batteries, fuel cells) are becoming increasingly critical \cite{report:IEA}. These devices are often fabricated from components prepared as slurries combining carbon black, forming a conductive colloidal gel \cite{article:Helal2016,article:Narayanan2017}, and larger granular particles such as graphite or other active components \cite{article:Entwistle2022,article:Wei2015,article:Kim2022}. These additives improve conductivity \cite{article:Sullivan2022,article:Don2022,article:Phillips2017,review:Antunes2011} but also non-trivially alter the slurry rheology and processability \cite{article:Kim2022,article:King2008,article:Kwon2015}. 

Here we combine rheometry, electron microscopy and dielectric measurements to examine the mechanical and electrical properties of a model slurry consisting of graphite and carbon black in mineral oil. We show that the material is bi-stable, with shear acting as a switch between a high-conductivity solid-like state and a low-conductivity fluid-like state, similar to recently reported bi-stability in silica-based binary suspensions \cite{Jiang2022}. Our results show that slurry processing conditions could dramatically impact device performance, and should help guide efforts to optimise mixing and coating processes.

\section{Experimental}

\subsection{Suspension preparation}

The synthetic graphite (Timrex KS150) and carbon black (Ensaco 250G) were used as supplied  by Imerys Graphite \& Carbon (Bodio, Switzerland). 
The fraction of graphite powder passing through a $63 \ \si{\mu m}$ aperture in vibrational sieving was collected for the investigations. Optical microscopy gave  a count-based particle size distribution, from which we estimate $d_{\rm 90} \approx 58 \ \si{\mu m}$ and $d_{\rm 50} \approx 24 \ \si{\mu m}$, with $d_{\rm x} = \mathrm{y}$ meaning $\mathrm{x \ \%}$ of the imaged particles have a diameter $\leq \mathrm{y}$ \cite{bookchapter:Snow2020}. The assumption in our analysis that the particles are spherical is acceptable given that KS150 graphite has a relatively low aspect ratio. Heavy mineral oil (330760, viscosity $\eta_0 = \SI{0.16}{\pascal\second}$ at \SI{23}{\celsius}) was obtained from Merck.

Carbon black powder was mixed with mineral oil in a sealed \SI{250}{\milli\liter} bottle by hand shaking followed by sonication (Bransonic 5510E-MT, frequency \SI{42}{kHz}, \SI{135}{W}) for an hour. To prepare the binary suspensions, the gel was manually mixed with the graphite powder to obtain a visually homogeneous suspension. Samples were stored at room temperature and shaken and stirred vigorously before being used for rheometry.

\subsection{Rheometry}

Rheological measurements were conducted on a stress-controlled AR-G2 rheometer (TA Instruments) with a crosshatched parallel plate geometry (diameter \SI{40}{\milli\meter}, gap height \SI{800}{\micro\meter}) and temperature kept at \SI{23}{\celsius} using a Peltier plate. To erase loading memory, the carbon black gel and binary suspensions were subjected to a shear rate of $\dot{\gamma} = 1000 \ \si{s^{-1}}$ for up to \SI{1200}{s} until the viscosity reached a plateau, unless otherwise stated. We visually checked for, but never found, sample fracture.

\subsection{Cryogenic scanning electron microscopy (Cryo-SEM)}

The scanning electron microscope (SEM) was a Zeiss Crossbeam 550 FIB-SEM with a Quorum Technologies Ltd PP3010T cryogenic attachment. After rheological measurements, the upper geometry was lifted and a small amount of sample was carefully transferred and spaced between two rivets. The rivets were placed in a sample holder and plunge-frozen into slush nitrogen before being loaded into a Quorum chamber and freeze-fractured at \SI{-140}{\celsius}. Afterwards, a \SI{5}{min} sublimation at \SI{-90}{\celsius} removed any surface ice introduced during the freeze-fracture process. The uncoated samples were imaged with at \SI{1}{kV} acceleration voltage. Simultaneous imaging using a Zeiss 'InLens' secondary electron detector and a secondary electron secondary ion (SESI) detector allowed differentiation between topography and composition.

Monte Carlo simulations were performed with the software CASINO (v3.3.0.4) to estimate secondary electron yields from sample constituents under our experimental conditions (acceleration voltage \SI{1}{kV} and beam diameter \SI{1.5}{nm}). A density of \SI{2.22}{g \ cm^{-3}} was used for the graphite (from the manufacturer), and paraffins consisting of 14, 29 and 43 carbon atoms were used to model the mineral oil with density \SI{0.87}{g \ cm^{-3}}. 

\subsection{Dielectric spectroscopy}

Dielectric spectroscopy measurements were performed in an HR-20 rheometer (TA Instruments) equipped with a smooth parallel plate geometry (diameter \SI{25}{\milli\meter}, gap height \SI{800}{\micro\meter}). TRIOS software (TA Instruments) provided control of a Keysight E4980A LCR impedance analyser. Short- and open-circuit measurements were conducted to correct the raw data,  collected in logarithmically-spaced frequency sweeps from \SI{20}{Hz} to \SI{2}{MHz} at an  amplitude of \SI{100}{mV}.

In the main text, we consider only the real part of the complex conductivity, $\mathrm{Re} \left(\sigma^*\right) = \sigma' = \omega \epsilon_r'' \epsilon_0$, with $\omega=2\pi f$ the frequency of the applied field, $\epsilon_r''$ the imaginary part of the complex permittivity, and $\epsilon_0$ the permittivity of free space. Data for $\sigma'\left(f \right)$ and $\epsilon_{\rm r}''\left(f \right)$ are given in the supplementary information (SI) (Figure S1). To obtain the low-frequency (dc) conductivity, $\sigma_{\rm dc}$, we fit the conductivity spectra by $\epsilon_{\rm r}'' = \sigma_{\rm dc}/ \left(\omega  \epsilon_0\right)$ in the frequency range 20-1000 Hz and 20-250 Hz for the composite and gel, respectively, since $\epsilon_{\rm r}''\left(\omega \right)$ exhibited a slope of -1 in this frequency range \cite{bookchapter:Schonhals2003}. At least two spectra were collected for each sample.

\section{Results and discussion}

\begin{figure}[htbp]
    \centering
    \includegraphics[width=.4\textwidth]{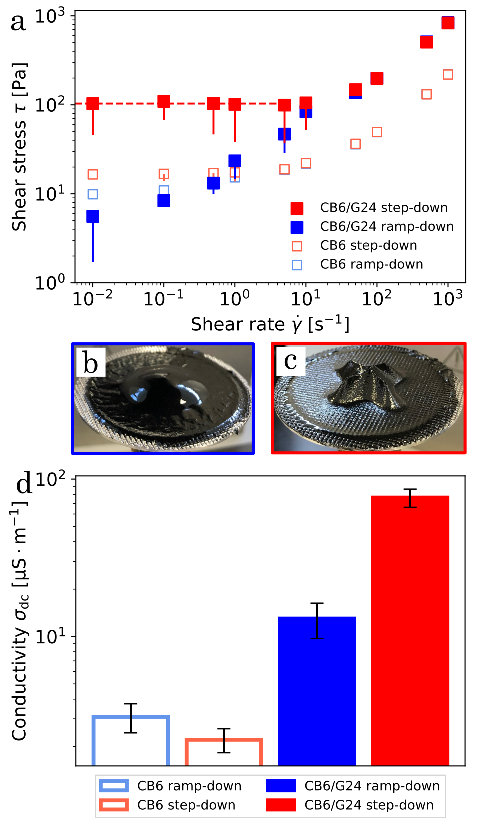}
        \caption{(a) Shear stress, $\tau$, as a function of shear rate, $\dot{\gamma}$, for \SI{6}{wt.\%} carbon black in oil (CB6) and \SI{6}{wt.\%} carbon black + \SI{24}{wt.\%} graphite in oil (CB6/G24). During step-down, the sample is rejuvenated at $\dot{\gamma}_{\rm rej} = 1000 \ \si{s^{-1}}$ for \SI{300}{s} between each $\dot{\gamma}$, and $\tau$ is recorded for \SI{300}{s}. Ramp-down is performed by progressively lowering $\dot{\gamma}$ with \SI{300}{s} at each step. Vertical bars indicate the range of $\tau$ measured after reaching the maximum stress (squares) at each $\dot{\gamma}$ when a steady state was not reached. Only bars larger than the marker size are shown. The horizontal dashed red line indicates the yield stress, $\tau_{\rm y}^{\rm H} = 103 \ \mathrm{Pa}$, of stepped-down CB6/G24. In (b,c), the composite after ramping down and quenching, respectively, is seen on the rheometer tool (diameter \SI{50}{mm}), prior to collecting material for cryo-SEM imaging. (d) dc electrical conductivity, $\sigma_{\rm dc}$, measured \SI{1000}{s} after the deformation protocol. The values of $\sigma_{\rm dc}$ $\left(10^{-5} - 10^{-4} \ \si{mS \cdot cm^{-1}} \right)$ are comparable to conductivities measured by Helal et al. \cite{article:Helal2016} in gels of CB with similar specific surface area to ours (\SI{65}{m^2/g} according to the manufacturer). Error bars denote $\pm$ one standard deviation based on measurements of at least two different samples.}
    \label{fig:results:rheometry:step-ramp-down}
\end{figure}

Individually, carbon black (CB) particles (diameter $\approx \SI{300}{\nano\meter}$ fused aggregates) in mineral oil form colloidal gels with finite yield stress for weight fractions above a few percent \cite{article:Trappe2000,article:Trappe2007,article:Gibaud2010}, while the larger graphite (G) particles (size $\approx \SI{30}{\micro\meter}$) in oil is an adhesive, non-Brownian suspension, also with a finite yield stress  \cite{Richards2021,article:Larsen2023}. So, we might expect the graphite to act simply as an `active filler' enhancing the gel yield stress \cite{article:Chen1999}. Instead, comparing the flow curves of a binary suspension (6 wt.\% CB, 24 wt.\% G) to a 6 wt.\% CB gel, we find a remarkable dependence on deformation protocol in the binary system that is absent in the CB-only gel, Figure \ref{fig:results:rheometry:step-ramp-down}a. When the shear stress, $\tau(\dot\gamma)$, is progressively ramped down, the binary suspension displays a weak yield stress ($\lesssim \SI{5}{\pascal}$) in the $\dot\gamma \to 0$ limit, {\it below} that of the un-filled CB-only gel ($\gtrsim \SI{10}{\pascal}$). If we instead rejuvenate the sample by sustained rapid shearing ($\dot{\gamma} = 1000 \ \si{s^{-1}}$ for \SI{300}{s}) between each applied shear rate, the yield stress of the binary suspension increases by over an order of magnitude to $\tau_{\rm y}^{\rm H} = \SI{103}{\pascal}$. This strong history dependence is evident when we inspect samples in the rheometer after measurement: a rejuvenated sample displays sharp peaks (Figure \ref{fig:results:rheometry:step-ramp-down}c) while a ramped-down sample appears liquid-like (Figure \ref{fig:results:rheometry:step-ramp-down}b).  

This protocol dependence in mechanical properties is mirrored in electrical properties. Measuring the conductivity at rest following shear cessation, we find a close to order of magnitude larger conductivity in the high-yield-stress composite relative to the ramped-down composite, Figure \ref{fig:results:rheometry:step-ramp-down}d. Additionally, the more than an order of magnitude conductivity increase in the high-yield-stress composite compared to the high-yield-stress gel demonstrates CB-G synergism in determining the electrical properties.

In contrast to the composites, the ramped down CB-only gel, although mechanically weaker, exhibits a marginally higher conductivity than the quenched sample. A similar shear-induced rise in conductivity in CB gels \cite{article:Helal2016} has been attributed to the formation of `log-rolling' vorticity-aligned flocs \cite{article:Osuji2008,article:Grenard2011} driven by plate confinement \cite{article:Varga2019}. Rheo-confocal images show similar structures in our CB-only gels, Figure S3, and we find a strain-dependent conductivity in these gels that correlates with the formation and breakup of these flocs, Figure S2. So, vorticity-aligned floc formation is probably also responsible for the observed slightly shear-dependent conductivity in our ramped-down CB-only gel.

These combined rheological and electrical results show that our CB/G composite switches between stable high-conductivity solid-like and low-conductivity liquid-like states depending on the shear history.  Similar memory and bistability was recently observed in composites of colloidal gels and granular suspensions of silica particles \cite{Jiang2022}, where the transition between solid-like and liquid-like states originated from the collapse of the gel network into compact, disjoint blobs at moderate applied stress. We use cryogenic scanning electron microscopy (cryo-SEM) to visualise the microstructure of our model composite slurry to explore if a similar mechanism exists in our system. In secondary-electron images, the conductive filler particles (both large graphite flakes and small carbon black aggregates) appear bright  against the insulating oil. This is expected for our low acceleration voltage and consistent with Monte Carlo simulations comparing the secondary electron yield $\left(\delta \right)$ from graphite ($\delta\sim0.9$) versus various paraffins (to simulate mineral oil, $\delta\sim0.3$). Note that some bright, narrow lines in our micrographs (FL) are fracture lines created during sample preparation. These are clearly seen in secondary electron secondary ion (SESI) images (Figure S4) that highlight topography, and should not be confused with edges of graphite flakes.

\begin{figure}[htbp]
\centering
\includegraphics[width=\textwidth]{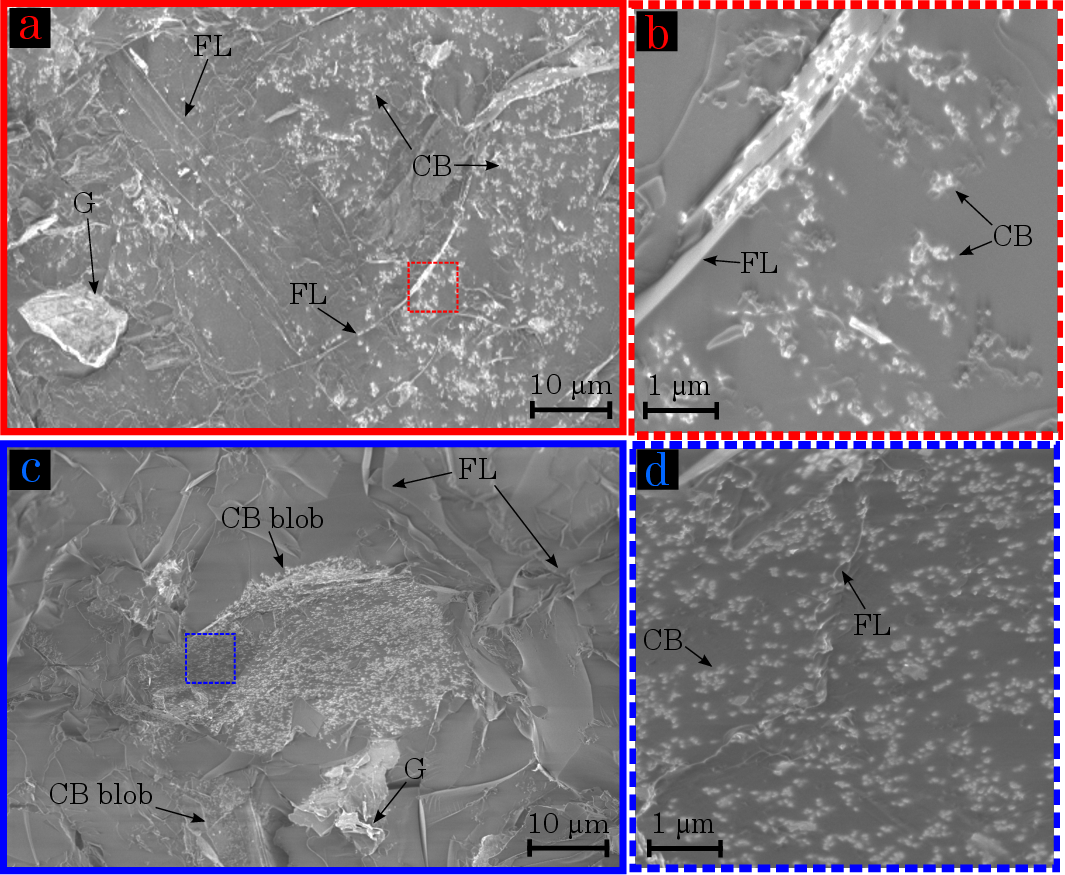}
\caption{Cryo-SEM images of our carbon black/graphite composite suspension after quenching to \SI{70}{Pa} $\left(< \tau_{\rm y}\right)$ for \SI{200}{s} following high-shear rejuvenation at $\dot\gamma=\SI{1000}{\per\second}$ (a,b). In (a), graphite (G) is seen as large (here $\sim 10 \ \si{\mu m}$), bright flakes, while the bright continuous, narrow lines are fracture lines (FL), not to be confused with graphite. Image (b), corresponding to the highlighted region in (a), shows individual carbon black (CB) aggregates, appearing as small bright particles, dispersed throughout the matrix. Images (c,d) show the CB/G suspension following the shear ramp-down protocol described in Figure \ref{fig:results:rheometry:step-ramp-down}a. Image (d) is a closer view at the region highlighted in (c), clearly showing the dense structure of CB blobs.}
\label{fig:cryo-SEM:1}
\end{figure}

\begin{figure}[htbp]
\centering
\includegraphics[width=\textwidth]{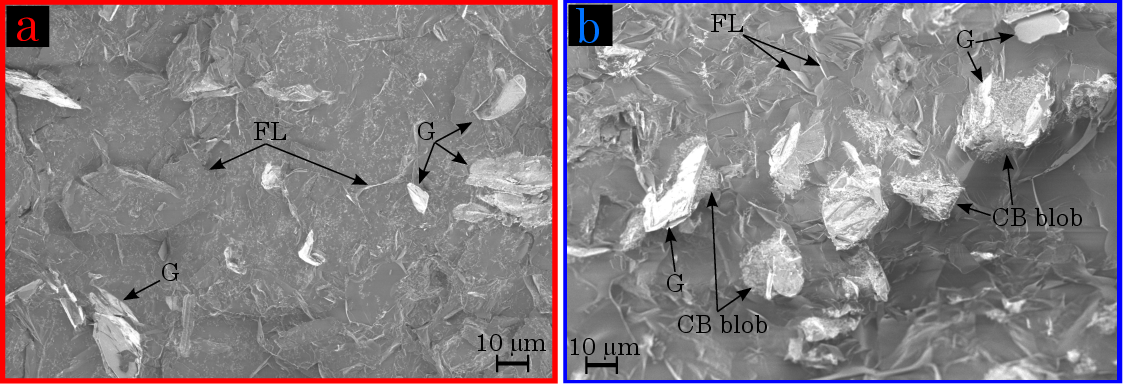}
\caption{Large area cryo-SEM images of our CB/G composites (a) after quenching to \SI{70}{Pa} $\left(< \tau_{\rm y}\right)$ for \SI{200}{s} following high-shear rejuvenation at $\dot\gamma=\SI{1000}{\per\second}$, and (b) after the ramp-down protocol. Examples of carbon black (CB) blobs, graphite (G) particles and fracture lines (FL) are indicated by arrows.}
\label{fig:cryo-SEM:2}
\end{figure}

In the high-yield-stress state produced by quenching from high-shear, Figure \ref{fig:cryo-SEM:1}a,b, irregular CB aggregates are evident throughout the sample, well dispersed between the flake-like graphite particles. In contrast, in the liquid-like state following shear ramp down, Figure \ref{fig:cryo-SEM:1}c,d, we find large, isolated blobs $\left(> 10 \ \si{\mu m} \right)$ of CB aggregates, with the surrounding fluid largely devoid of CB particles.  These CB blobs are notably denser than the well-dispersed CB particles in the solid-like state, compare Figures \ref{fig:cryo-SEM:1}b with \ref{fig:cryo-SEM:1}d. Large area images, Figure \ref{fig:cryo-SEM:2} (and Figures S5 and S6 in the SI), confirm the generality of these features.

The collapse of the CB gel phase into compact, disjoint blobs explains the decrease in conductivity and the disappearance of the yield stress following shear ramp down. The microstructure in our fluid-like state differs somewhat from that in the fluid-like states in previously-described silica-based composites~\cite{Jiang2022}. We find dense aggregations of CB particles generally forming around the larger graphite particles, while the small-particle blobs in the silica-based composites were generally disjoint form the larger particles. This suggests significant attraction between the large and small particles in our composites. This is perhaps unsurprising. The interactions in the previously-studied model binary silica suspensions were designed to be non-attractive between the small colloids and large grains. In our case, we have clear evidence of attractive interactions in pure CB and G suspensions. Since both CB and G are forms of carbon, we may then also expect attraction between CB and G particles. The resulting close proximity of CB blobs and G flakes allow the latter to act as bridges between the former, Figures \ref{fig:cryo-SEM:1}c and \ref{fig:cryo-SEM:2}b, which, together with the formation of densely packed blobs with a low intrinsic resistance \cite{article:Narayanan2017}, may explain why the contrast in conductivity between the two states in our system is less dramatic than the contrast in the yield stress.

Our results so far show that our CB/G slurry exhibits a shear induced solid/liquid transition similar to that previously observed in a mixture of attractive colloids and repulsive granular particles~\cite{Jiang2022}. In the previous system, the solid/liquid transition was controlled by two stress scales: the yield stress of the homogeneous solid-like state, $\tau_{\rm y}^{\rm H}$, and a characteristic stress for breakup of the large blobs in the liquid-like state, $\tau_{\rm B}$. 

To see if this is also the case in our system, we use creep tests to characterise the yield stress of our composites, varying the preshear stress $\tau_{\rm pre}^{\rm ss}$. In these tests, we measure the time-dependent shear rate $\dot\gamma(t)$ for a sequence of increasing fixed stresses $\tau$, identifying the $\tau_{\rm y}$ from the onset of steady flow, $\dot\gamma(t)\approx{\rm const.}$, in the long-time limit. Prior to each test, the composite is prepared in either a heterogenous or homogenous state, followed by extended shear at the given preshear stress $\tau_{\rm pre}^{\rm ss}$.

\begin{figure}[htbp]
\centering
\includegraphics[width=.5\textwidth]{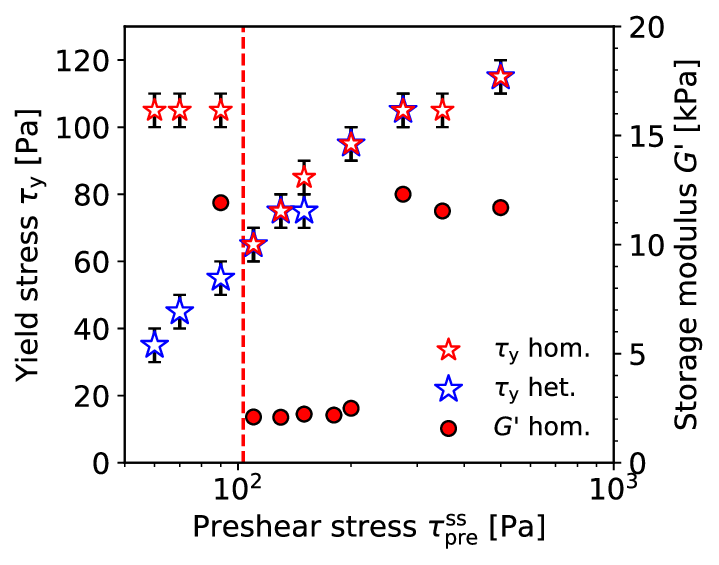}
\caption{Yield stresses (open stars) from creep tests (error bars indicate the finite stress step size) on CB6/G24 after steady shear preshearing at the indicated preshear stresses, $\tau_{\rm pre}^{\rm ss}$. The sample was either initially in a homogeneous state (hom.), obtained by pre-preshearing at $\dot{\gamma} = 1000 \ \si{s^{-1}}$ for \SI{300}{s}, or a heterogeneous state (het.) by pre-preshearing at $\dot{\gamma} = 10 \ \si{s^{-1}}$ for \SI{1000}{s}. During subsequent preshearing, the accumulated strain at each $\tau_{\rm pre}^{\rm ss}$ was $\gamma_{\rm total} = 20,000$, except at $\tau_{\rm pre}^{\rm ss} \leq 90 \ \si{Pa}$ for CB6/G24 Hom.~as the sample did not flow below its yield stress, $\tau_{\rm y}^{\rm H}=103 \ \mathrm{Pa}$ estimated from the flow curve in Figure \ref{fig:results:rheometry:step-ramp-down}a, and indicated by the vertical dashed red line, so a \SI{200}{s} preshear was applied instead. For the composite initially in the homogeneous state, the storage modulus $G'$ (solid circles), measured in the linear viscoelastic region (LVR), strain amplitude $\gamma_0^{\rm osc} = 0.001$ and frequency \SI{10}{rad \ s^{-1}}, was obtained by appending a \SI{200}{s} shear step at \SI{90}{Pa} after shearing at the indicated $\tau_{\rm pre}^{\rm ss}$. Then, $G'$ was noted after \SI{100}{s} in the LVR.}
\label{fig:4}
\end{figure}

At high preshear stresses, Figure \ref{fig:4}, we obtain the yield stress of the homogeneous state, $\tau_{\rm y}\approx\SI{100}{\pascal}\approx\tau_{\rm y}^{\rm H}$, irrespective of the initial state prior to applying $\tau_{\rm pre}^{\rm ss}$.  For preshear stresses $\tau_{\rm pre}^{\rm ss}\lesssim\SI{200}{\pascal}$, the $\tau_{\rm y}$ steadily decreases with decreasing $\tau_{\rm pre}^{\rm ss}$, roughly halving for  $\tau_{\rm pre}^{\rm ss}$ slightly above the homogeneous yield stress.  For $\tau_{\rm pre}^{\rm ss}>\tau_{\rm y}^{\rm H}$, the yield stress remains independent of the initial state prior to the preshear step.  However, preshearing below $\tau_{\rm y}^{\rm H}$ reveals a bistable regime. Initially homogeneous solid-like samples retain their yield stress, so that $\tau_{\rm y}\approx \tau_{\rm y}^{\rm H}\approx{\rm constant}$ for $\tau_{\rm pre}^{\rm ss}<\tau_{\rm y}^{\rm H}$, while the yield stress in the initially heterogeneous sample continues to decrease with decreasing $\tau_{\rm pre}^{\rm ss}$. 

Thus, our graphite/carbon black composite exhibits similar bi-stable rheology to that observed in the silica composites~\cite{Jiang2022}. For preshear stresses above the homogeneous yield stress, $\tau_{\rm pre}^{\rm ss}>\tau_{\rm y}^{\rm H}$, the sample response is history-independent but depends on $\tau_{\rm pre}^{\rm ss}$, with $\tau_{\rm y}\approx \tau_{\rm y}^{\rm H}={\rm constant}$ for high preshear stresses but then weakening over a range of moderate preshear stresses, indicating the transition from the homogeneous to the heterogeneous state. However, the yield stress of our composite remains finite in the heterogeneous state, which might be expected given the large-small attraction in our composites. With the large-small attraction absent, or at least dramatically reduced in the composites previously studied~\cite{Jiang2022}, the percolated small-particle network that underpins the solid-like response is suddenly lost in the heterogenous blob-like state. However, in our graphite/carbon black composite both graphite-graphite and graphite-blob contacts enable the retention of a connected, albeit weaker, structure in the heterogenous state. In this weaker heterogenous state, the composite's yield stress should depend on the shape and structure of the blobs, so that $\tau_{\rm y}$ varies with the preshear stress in this regime.    Thus, while the bi-stable regime below $\tau_{\rm y}^{\rm H}$ is clear in both systems,  the transition between liquid and solid states at higher stresses is less clear in our CB/G composites when characterised by the yield stress.

A clearer mapping onto previously-found behaviour~\cite{Jiang2022} results when we measure the small-amplitude storage modulus ($G^\prime$) of an initially solid-like composite after appending a low-stress creep step (shearing at $\SI{90}{\pascal}< \tau_{\rm y}^{\rm H}$) between our pre-shear step at $\tau_{\rm pre}^{\rm ss}$ and our oscillatory measurement, Figure \ref{fig:4}a (filled symbols). Now, we find a sharp jump in elastic modulus with $G'\gtrsim\SI{e4}{\pascal}$ above $\tau_{\rm pre}^{\rm ss} \approx \SI{200}{\pascal}$ and below $\tau_{\rm y}^{\rm H}=\SI{103}{\pascal}$. This is consistent with the initial drop in $\tau_{\rm y}$ at intermediate pre-shear stresses, suggesting an estimate for the blob breakup stress $\tau_{\rm b}\approx\SI{200}{\pascal}$. While we do not fully understand why the transition appears sharper when measuring the elastic modulus compared to the yield stress, we might expect adhesive large particle contacts to be broken under small amplitude oscillations \cite{Richards2020CSadh}, so that $G'$ provides a more direct probe of the small particle network. We thus recover the essential rheological features of the granular/gel composites described previously~\cite{Jiang2022}, with the solid-liquid transition controlled by two stress scales: the yield stress of the homogeneous composite $\tau_{\rm y}^{\rm H}$ and a characteristic blob breakup stress $\tau_{\rm b}$.

\section{Summary and conclusions}

We have shown that binary suspensions of graphite and carbon-black particles display a similar mechanical bistability as mixtures of granular and colloidal silica particles~\cite{Jiang2022}. Prolonged shear of the mixture at high enough applied stress followed by abrupt cessation of shear produces a high-yield-stress solid state -- a gel of the smaller particles in which the larger particles are dispersed homogeneously. However, a gradual ramping down of the shear rate produces either a liquid state (silica mixtures) or a low-yield-stress solid (the present system) in which the smaller particles are aggregated into discrete blobs. Such structure-rheology correlations in cathode slurries have been reported before~\cite{article:Park2022,Komoda2022}, with some evidence for history dependence~\cite{article:Park2022}, so that bistability may be generic to binary suspensions of large grains dispersed in colloidal gels. 

In our case, we find that the mechanically-stronger solid also has higher electrical conductivity. This finding in our model system can be directly compared with a similar correlation between higher conductivities and discharge capacities in more elastic anode slurries for Li-ion batteries~\cite{article:Sullivan2022}. Our cryo-SEM images provide a direct, particle-level explanation of the link between mechanical strength and electrical conductivity: the higher-performance samples are produced using a shearing protocol that completely homogenises the system, while lower-performance samples are associated with a highly-heterogeneous distribution of CB particles. A recent study using SEM and energy-dispersive X ray spectroscopy~\cite{article:Saraka2020} points to a similar conclusion for the relation between microstructure and performance in dried slurry coatings used as electrodes in Li-ion batteries. This study does not, however, discuss the relative importance of slurry processing and the subsequent processes of coating and drying in determining the microstructure. Our results suggest that slurry processing certainly plays a key role. Future work may therefore fruitfully examine how the bistability of the slurry may also influence the outcome of the coating and drying process. 
\section*{Declaration of competing interests}

The authors declare that they have no known competing financial
interests or personal relationships that could have appeared to influence
the work reported in this paper.

\section*{Data availability}

The presented data can be accessed via \url{https://doi.org/10.7488/ds/7538}.

\section*{Acknowledgements}

Graphite and carbon black powders were kindly provided by Imerys Graphite \& Carbon.
Financial support by Advent Technologies A/S and a grant from the Industrial PhD programme, Innovation Fund Denmark, project 8053-00063B is gratefully acknowledged.

\newpage


\printbibliography

\end{document}